\documentclass{article}

\usepackage{arxiv}
\pdfoutput=1

\usepackage[utf8]{inputenc} 
\usepackage[T1]{fontenc}    
\usepackage{hyperref}       
\usepackage{url}            
\usepackage{booktabs}       
\usepackage{amsfonts}       
\usepackage{nicefrac}       
\usepackage{microtype}      
\usepackage{cleveref}       
\usepackage{lipsum}         
\usepackage{graphicx}
\usepackage[numbers]{natbib}
\usepackage{doi}

\title{Localized polariton states in a photonic crystal intercalated by a transition metal dichalcogenide monolayer}

\date{}

\newif\ifuniqueAffiliation

\ifuniqueAffiliation 
\author{ \href{https://orcid.org/0000-0000-0000-0000}{\includegraphics[scale=0.06]{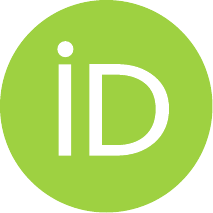}\hspace{1mm}Yu.~V.~Bludov}\thanks{Use footnote for providing further
		information about author (webpage, alternative
		address)---\emph{not} for acknowledging funding agencies.} \\
	Department of Physics, Center of Physics, and QuantaLab,\\
	University of Minho, Campus of Gualtar,\\
	4710-057, Braga, Portugal \\
	\texttt{hippo@cs.cranberry-lemon.edu} \\
	\And
	\href{https://orcid.org/0000-0000-0000-0000}{\includegraphics[scale=0.06]{orcid.pdf}\hspace{1mm}Elias D.~Striatum} \\
	Department of Electrical Engineering\\
	Mount-Sheikh University\\
	Santa Narimana, Levand \\
	\texttt{stariate@ee.mount-sheikh.edu} \\
}
\else
\usepackage{authblk}

\newbox{\orcid}\sbox{\orcid}{\includegraphics[scale=0.06]{orcid.pdf}} 
\author[1]{%
	\href{https://orcid.org/0000-0001-9648-1459}{\usebox{\orcid}\hspace{1mm}Yu.~V.~Bludov\thanks{\texttt{bludov@fisica.uminho.pt}}}%
}
\author[2]{%
	C. Fernandes%
}
\author[1,2]{%
	\href{https://orcid.org/0000-0002-7928-8005}{\usebox{\orcid}\hspace{1mm}N.~M.~R.~Peres}%
}
\author[1,2]{%
	\href{https://orcid.org/0000-0002-7928-8005}{\usebox{\orcid}\hspace{1mm}M.~I.~Vasilevskiy}%
}
\affil[1]{Department of Physics, Center of Physics, and QuantaLab, University
	of Minho, Campus of Gualtar, 4710-057, Braga, Portugal}
\affil[2]{International Iberian Nanotechnology Laboratory
	(INL), Av. Mestre Jos\'{e} Veiga, 4715-330 Braga, Portugal}
\fi

\renewcommand{\shorttitle}{Localized polariton states in a photonic crystal}

\hypersetup{
pdftitle={Localized polariton states in a photonic crystal intercalated by a transition metal dichalcogenide monolayer},
pdfauthor={Yu.~V.~Bludov, C.~Fernandes, N.~M.~R.~Peres, M.~I.~Vasilevskiy},
}

\begin{document}
\maketitle

\begin{abstract}
Beyond the extensively studied microcavity polaritons, which are coupled modes  of semiconductor excitons and microcavity photons, nearly 2D semiconductors placed in a suitable environment can support spatially localized exciton-polariton modes. We demonstrate theoretically that two distinct types of such modes can exist in a photonic crystal with an embedded transition metal dichalcogenide (TMD) monolayer and derive an equation that determines their dispersion relations. The localized modes of two types occur in the zeroth- and first-order stop-bands of the crystal, respectively, and have substantially different properties. The latter type of the localized modes, which appear inside the light cone, can be described as a result of coupling of the TMD exciton and an optical Tamm state of the TMD-intercalated photonic crystal. We suggest an experiment for detecting these modes and simulate it numerically. 
\end{abstract}


\section{Introduction}

Since the discovery of two-dimensional (2D) one-atomically thick material
graphene in 2004 \citep{gr-Novoselov2004-science}, in the area of
electronics it was great demand for the similar materials, but with
bandgaps in their electronic spectrum. After several years, several 2D semiconductors were found, such as the transition metal dichalcogenide (TMD) family \cite{TMD-Mak2010-prl,TMD-Splendiani2010-nl}
and phosphorene \cite{phosphorene-Arra2019-prb}.
In particular, TMDs possess bandgaps of width corresponding to the
optical range of wavelengths and they behave as 2D semiconductors.
The 2D nature of the TMDs leads to reduced dielectric screening and, consequently, strong Coulomb interaction between electrons and holes, which results in the formation of tightly bound excitons \cite{TMD-exciton-He2014-prl}. The excitonic luminescence of these materials is of practical interest and can be enhanced and even the lasing regime can be achieved by incorporating the 2D layer into an appropriate photonic structure \cite{Ye2015,tmd-pc-Liu_2017-scirep,Chen2020}. 

\begin{figure}[t!]
	\centering
	\fbox{\includegraphics[width=10cm]{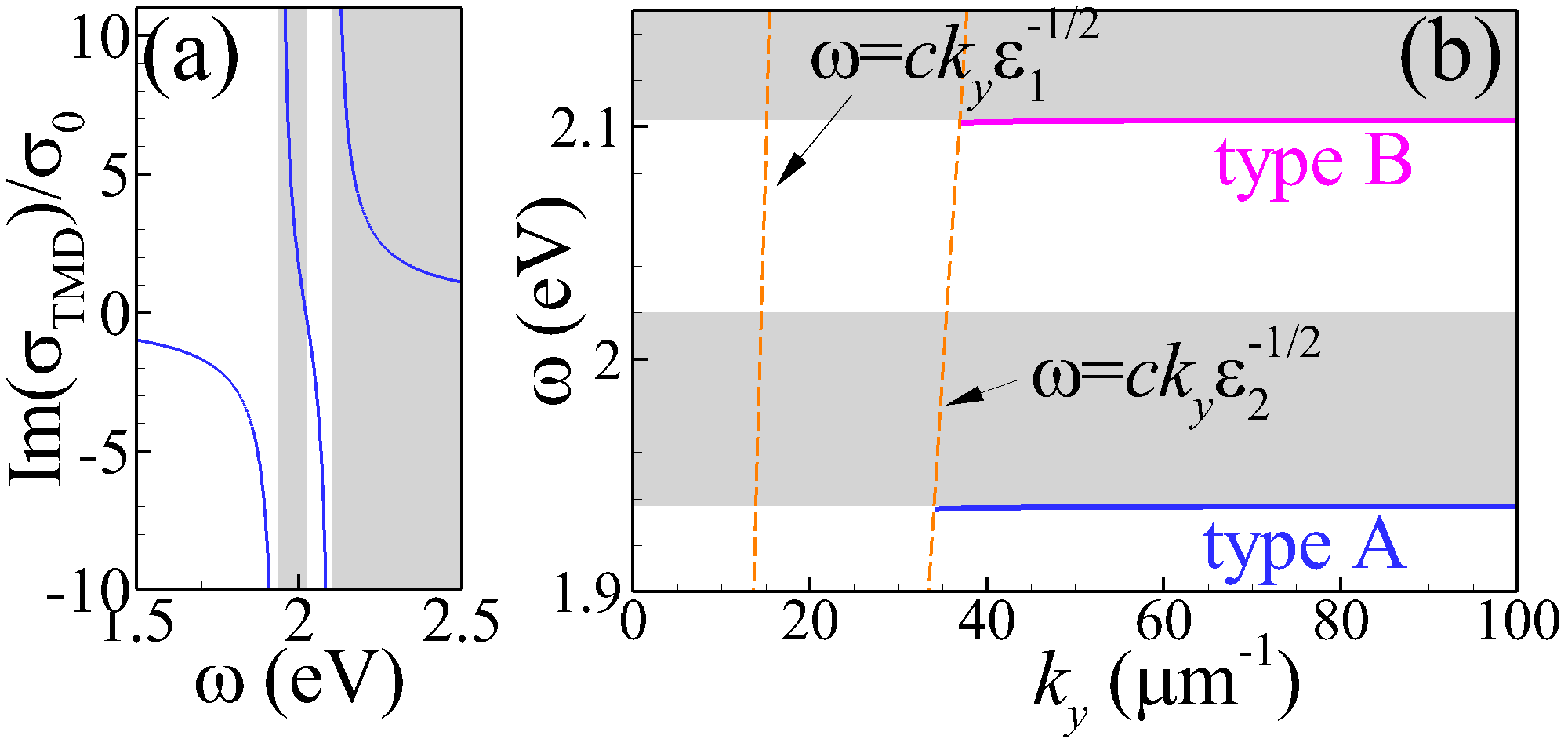}}
	
	\caption{(a) Frequency dependence of the imaginary part of the $\mathrm{MoS}{}_{2}$ optical 
		conductivity for $\gamma _A=\gamma _B=0$; (b) Dispersion relation of surface EPs (solid blue and pink lines), supported by a single $\mathrm{MoS}{}_{2}$ layer located at the interface between two dielectrics with $\varepsilon_{1}=2$
		and $\varepsilon_{2}=12$. In both panels the frequency regions
		with positive Re$\sigma$ are shadowed. In panel (b) the light lines $\omega=ck_{y}/\sqrt{\varepsilon_{m}}\quad\left(m=1,2\right)$
		are depicted by orange dashes. Here and throughout the paper
		we consider the following parameters for the $\mathrm{MoS}{}_{2}$
		conductivity (\ref{eq:conductivity}): $P_{A}=0.2530\thinspace\mathrm{eV}$,
		$\omega_{A}=1.93715\thinspace\mathrm{eV}$, $P_{B}=0.2517\thinspace\mathrm{eV}$,
		$\omega_{B}=2.10327\thinspace\mathrm{eV}$.}
	
	\label{fig:disp-rel-semiinfinite}
\end{figure}

Generally, the optical spectra of TMDs are characterized
by the presence of two excitonic transitions, referred to as type A and type B \cite{Wang2018}. If TMD layer is embedded into a microcavity (MC),  these excitons can effectively couple to MC photons, forming MC exciton-polaritons (EPs) \cite{TMD-pc-excit-pol-Liu2015-natphot,TMD-cav-excit-pol-Dufferwiel2015-ncomm,Flatten2016,Lundt2016,TMD-cav-excit-pol-Vasilevskiy2015-prb,Zhang2018,TMD-excit-pol-Hu2019-prb}. This exciton-light
coupling can be described by the 2D optical conductivity of a TMD layer, 
\begin{eqnarray}
\sigma_{TMD}\left(\omega\right)=\sigma_{0}\sum_{j=A,B}\frac{P_{j}}{\gamma _j+i\omega_{j}-i\omega}\,,\label{eq:conductivity}
\end{eqnarray}
which takes into account the aforementioned excitonic transitions \cite {TMD-cav-excit-pol-Vasilevskiy2015-prb}. Its real part exhibits sharp peaks at the excitonic transition
frequencies, $\omega_{A}$ and $\omega_{B}$ ($\omega_{A}<\omega_{B}$) and in Eq.\,(\ref{eq:conductivity}) $P_{A\,(B)}$ stands
for longitudinal-transverse splitting of exciton
A (B). Also, $\gamma_{A\,(B)}$ are damping parameters and $\sigma_{0}=e^{2}/\left(4\hbar\right)$ the quantum of conductivity. The imaginary part of the conductivity changes sign in the vicinity of the excitonic transition frequencies {[}see Fig.\,\ref{fig:disp-rel-semiinfinite}(a){]}. As a result, the TMD is characterized
by a negative imaginary part of the conductivity in two frequency
ranges, $\omega<\omega_{A}$ and $\omega_{*}<\omega<\omega_{B}$
(white regions in Fig.\,\ref{fig:disp-rel-semiinfinite}), while
${\rm Im}(\sigma_{TMD})$ is positive for $\omega_{A}<\omega<\omega_{*}$ and $\omega>\omega_{B}$ (grey regions in Fig.\,\ref{fig:disp-rel-semiinfinite}). Here $\omega_{*}=\left(P_{A}\omega_{B}+P_{B}\omega_{A}\right)/\left(P_{A}+P_{B}\right)$.

If TMD is cladded by two semi-infinite dielectric media with
dielectric constants $\varepsilon_{1}$ and $\varepsilon_{2}$, it is able to sustain {\it surface} EPs with a dispersion relation, $\omega(k_y)$, determined by the equation \cite {TMD-cav-excit-pol-Vasilevskiy2015-prb} 
\begin{eqnarray}
k_{z}^{(1)}+k_{z}^{(2)}+\frac{4\pi\omega}{c^{2}}\sigma_{TMD}\left(\omega\right)=0\,,
\label{disprel1}
\end{eqnarray}
where $k_{z}^{(m)}=\sqrt{\kappa^{2}\varepsilon_{m}-k_{y}^{2}}$ and
$k_{y}$ are out-of-plane and in-plane components of the wave vector, respectively, 
in the medium with dielectric constant $\varepsilon_{m}$ $(m=1,2)$, 
$\kappa=\omega/c$, and $c$ is the velocity of light in vacuum. Neglecting damping, $k_z^{(1)}$ and $k_z^{(2)}$ are purely imaginary for surface EPs, in contrast with the MC polaritons. As an
example, the dispersion of surface EPs in a single layer of $\mathrm{MoS}{}_{2}$
is depicted in Fig.\,\ref{fig:disp-rel-semiinfinite}(b). 
There are two such modes {[}type A and type B, depicted in Fig.\,\ref{fig:disp-rel-semiinfinite}(b) by solid blue and pink lines, correspondingly{]}, which exist in
the frequency ranges where imaginary part of TMD conductivity is
negative (white domains in Fig.\,\ref{fig:disp-rel-semiinfinite}), as it happens for s-polarized plasmon-polaritons in graphene \cite{spp-graphene-TE-Mikhailov2007}.
Notice that both A and B-type surface EP modes occur outside of the light cone, their dispersion curves bifurcate from the less steep light line, $\omega=ck_{y}/\sqrt{\mathrm{max}\left(\varepsilon_{1},\varepsilon_{2}\right)}$, and asymptotically approach the exciton transition frequencies, $\omega_{A}$ and $\omega_{B}$, at large values of $k_{y}$. Alike other evanescent waves, surface EPs cannot be excited  directly by external propagating electromagnetic (EM) waves.

As mentioned above, the efficiency of the exciton-light coupling can be enhanced considerably if the TMD layer is placed in an optical resonance system, such as Fabry-Perot \cite{TMD-cav-excit-pol-Dufferwiel2015-ncomm,TMD-cav-excit-pol-Vasilevskiy2015-prb}, micropillar \cite{TMD-pc-excit-pol-Liu2015-natphot,NGomes_JAP2020} or Tamm-plasmon \cite{Flatten2016,Lundt2016} microcavity, on top of a specially prepared metasurface \cite{Zhang2018,Chen2020} or just near a plasmonic surface \cite{TMD-spp-Goncalves2018-prb}. In this way, the strong coupling regime can be achieved, which offers a range of potential applications (see references in the recent papers \cite {Chen2020,NGomes_JAP2020}). The presence of a MC confining the light allows for the existence of the "{\it bulk}"-type MC polaritons with real $k_z^{(1)}$, $k_z^{(2)}$ \cite{Kavokin_MCs}, which coexist with the \emph {surface} EPs \cite{TMD-cav-excit-pol-Vasilevskiy2015-prb}.

However, it may be not a MC but rather a heterostructure formed by two finite PCs (or Bragg reflectors, BR, in other words) where the latter type of modes exist. The existence of lossless electromagnetic (EM)  interface modes at the boundary of such a heterostructure was demonstrated in Ref. \cite{Kavokin2005}, where they were named optical Tamm states; later the term passed to designate mostly localized (in one direction) EM modes formed in the gap between a BR and a metal surface, coupled to metal plasmons \cite{TammPlasmonTP,Gazzano2012,sasin_2010}.
In this paper we demonstrate that a TMD monolayer embedded in a photonic crystal (PC) is able to sustain localized EP eigenstates, similar to other types of "defects" in PCs, which break the translational symmetry along the PC axis \cite{def-Yablonovich1991-prl,def-Villeneuve1996-prb,def-waveguide-Notomi2001-prl,def-cavity-Kuramochi2006-apl,def-emit-Noda2000-nature,def-nonl-Valligatla2015-optmat}. These states are characterized by a well-defined real in-plane component of the $k$-vector, i.e. they correspond to evanescent waves coupled to the TMD excitons. Such modes can be excited directly by external light if the number of periods in the PC is not too large. Moreover, using oblique incidence of light and measuring absorbance, one can probe the dispersion relation of these modes as demonstrated by our numerical simulation.

\section{Electromagnetic eigenmodes}
\begin{figure}[t!]
	\centering
	\fbox{
		\includegraphics[width=10cm]{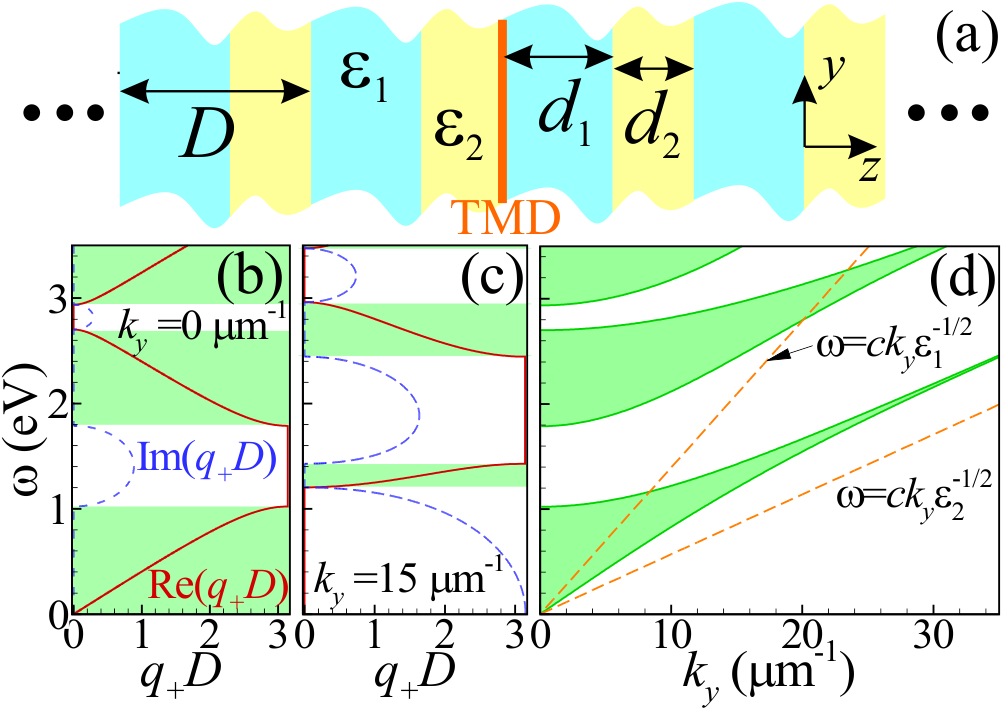}
	}
	\caption{(a) Schematic representation of a photonic crystal with embedded TMD layer;
		(b,c) Relation between the frequency $\omega$ and the Bloch wavevector
		of a forward-propagating wave, $q_{+}$, for two values of $k_{y}$. Red solid (dashed blue) lines correspond
		to real (imaginary) parts of $q_{+}$. (d) Frequencies of allowed
		bands' edges (solid green lines) plotted against in-plane wavevector, 
		$k_{y}$ and two light lines depicted by orange dashes. 
		The domains corresponding to allowed bands are shadowed green. The following PC parameters were used: $\varepsilon_{1}=2$,
		$d_{1}=140\thinspace$nm, $\varepsilon_{2}=12$, $d_{2}=70\thinspace$nm.}
	\label{fig:infinite_PC_TE}
\end{figure}

Let us consider a photonic crystal with a period $D$, composed
of two alternating dielectric layers {[}along $z$-axis, see Fig.\,\ref{fig:infinite_PC_TE}(a){]}, a layer with the dielectric constant $\varepsilon_{1}$ and thickness $d_{1}$
(which occupies spatial domains $nD<z<nD+d_{1}$) and a layer with dielectric
constant $\varepsilon_{2}$ and thickness $d_{2}=D-d_{1}$ {[}arranged
at spatial domains $nD+d_{1}<z<(n+1)D${]}. Here $n$ stands for the
PC cell's number. We also suppose that the TMD layer is placed at the boundary between two dielectrics (plane $z=0$). If the EM field is uniform in $x$-direction ($\partial/\partial x\equiv0$), Maxwell's equations for a TE-polarized wave (containing components
$E_{x}$, $H_{y},$ and $H_{z}$) read:
\begin{eqnarray}
& ik_{y}H_{z}^{(m,n)}-\frac{\partial H_{y}^{(m,n)}}{\partial z}=-i\kappa\varepsilon_{m}E_{x}^{(m,n)},\label{eq:Maxwell-TE-1}\\
& \frac{\partial E_{x}^{(m,n)}}{\partial z}=i\kappa H_{y}^{(m,n)},\quad
-ik_{y}E_{x}^{(m,n)}=i\kappa H_{z}^{(m,n)}\,.\label{eq:Maxwell-TE-3}
\end{eqnarray}
Here the spatial and temporal dependence of the EM field of the form $\propto\exp\left(ik_{y}y-i\omega t\right)$ is assumed. 
Notice that 
the composite superscript $\left(m,n\right)$ in Eqs.\,(\ref{eq:Maxwell-TE-1})--(\ref{eq:Maxwell-TE-3}) is prescribed to the EM
field components defined in the $n$-th unit cell of the PC, in its part filled with
the dielectric $\varepsilon_{m}$. After solving Maxwell's equations in each slab and applying boundary conditions (see Supplemental Document for details), it is possible to relate the tangential components of the field across the PC period via the unit cell transfer matrix $\hat{T}_{12}$,
	i.e.,
\begin{eqnarray}
\left(\begin{array}{c}
H_{y}^{(1,n+1)}\left(\left(n+1\right)D\right)\\
E_{x}^{(1,n+1)}\left(\left(n+1\right)D\right)
\end{array}\right)=\hat{T}_{12}\left(\begin{array}{c}
H_{y}^{(1,n)}\left(nD\right)\\
E_{x}^{(1,n)}\left(nD\right)
\end{array}\right).\label{eq:HE-period}
\end{eqnarray}

The dispersion relation of electromagnetic waves in a \emph {perfect infinite PC} can be obtained by applying the Bloch theorem. It can be represented
in terms of eigenvalues, $\lambda_{\pm}$, and eigenvectors,
$\left(h_{y}^{(\pm)}\quad e_{x}^{(\pm)}\right)^{T}$, of the matrix $\hat{T}_{12}$ (see Eq.\,(S2) of Supplemental Document). Namely, the following equation holds:
\begin{eqnarray}
\exp\left(iq_{\pm}D\right)=\lambda_{\pm}\,.
\label{disp_eq}
\end{eqnarray}
The eigenvalues, $\lambda_{+}$ and $\lambda_{-}$, determine
the dispersion relations for forward- and backward-propagating waves,
respectively. Moreover, as the matrix $\hat{T}_{12}$
is unimodular, the pair of eigenvalues possess the property $\lambda_{+}\lambda_{-}=1$,
which implies the relation $q_{-}=-q_{+}$.

\begin{figure}[t!]
	\centering
	\fbox{
		\includegraphics[width=10cm]{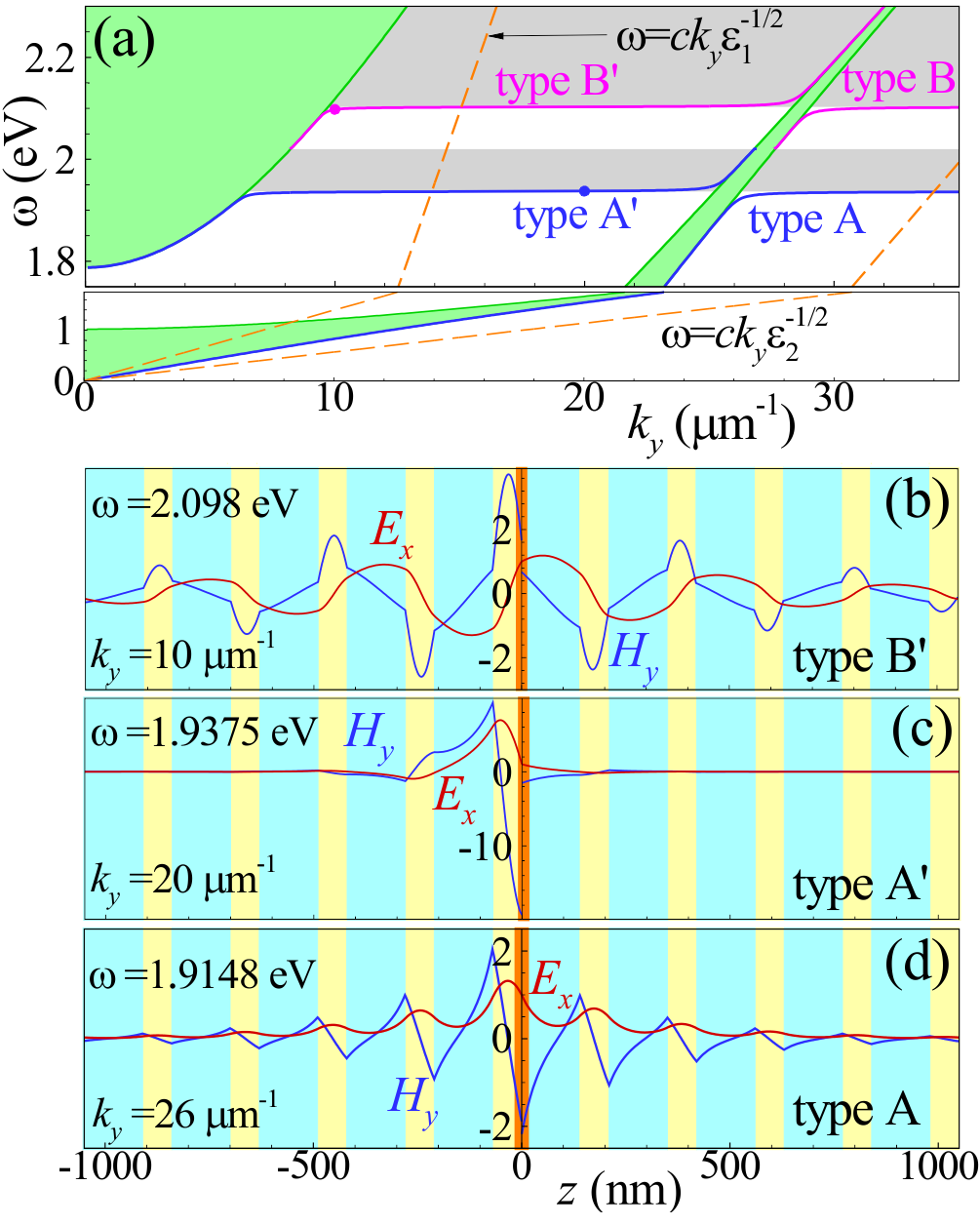}
	}
	\caption{(a) Dispersion curves (solid blue and pink lines) and (b)--(d)
		spatial profiles of EM field components $H_{y}(z)$ and
		$E_{x}(z)$ (depicted by blue and red lines, respectively) corresponding to the localized EP eigenstate in the PC with TMD. In (a), allowed bands, their edges and light
		lines are depicted as in Fig.\,\ref{fig:infinite_PC_TE}, while the
		regions of positive and negative ${\rm Im }\sigma_{TMD}$ are shown as in Fig.\,\ref{fig:disp-rel-semiinfinite}. Panels (b)--(d): spatial profiles of the EM field components corresponding to the parameters
		($\omega$ and $k_{y}$) marked by filled circles in panel (a). The A (B) and A' (B') modes are explained in the text.}
	\label{fig:loc-state-TE}
\end{figure}

The Bloch wavevector can be either real or imaginary (if damping is neglected), which depends on whether the chosen pair of $\omega$ and $k_{y}$ belongs to the allowed or
forbidden (also called stop-band) part of the PC's spectrum. Namely, inside the allowed bands {[}green-shadowed domains in Figs.\,\ref{fig:infinite_PC_TE}(b) and \ref{fig:infinite_PC_TE}(c){]}, the Bloch wavevector of forward-propagating wave is purely real and positive, i.e., $\mathrm{Re}\left(q_{+}\right)\ge0$, $\mathrm{Im}\left(q_{+}\right)=0$.
In contrast, inside the stop-bands {[}white domains in Figs.\,\ref{fig:infinite_PC_TE}(b)
and \ref{fig:infinite_PC_TE}(c){]}, the Bloch wavevector can be treated as a complex value with the real part $\mathrm{Re}\left(q_{+}\right)=0,\, \pi/D$ and a positive imaginary part
$\mathrm{Im}\left(q_{+}\right)>0$. The latter implies the evanescent character
of forward-propagating wave inside the stop-band and its decrease
in the positive direction of $z$-axis. Accordingly, $\mathrm{Im}\left(q_{-}\right)<0$ for a backward-propagating wave inside the stop-band, i.e. its amplitude
decreases in the negative direction of $z$-axis. 

Let us now turn to a \emph {PC intercalated by a TMD layer}. The fields in each part of it, at planes $z=nD$ with $n$ standing for an integer, either positive or negative, can be represented as follows (see Eq. (S3) of Supplemental Document):
\begin{eqnarray*}
	&\left(\begin{array}{c}
		H_{y}^{(2,n-1)}\left(nD\right)\\
		E_{x}^{(2,n-1)}\left(nD\right)
	\end{array}\right)=A_{-}\left(\begin{array}{c}
		h_{y}^{(-)}\\
		e_{x}^{(-)}
	\end{array}\right)\exp\left(iq_{-}nD\right),\thinspace n\le0,\\
	&\left(\begin{array}{c}
		H_{y}^{(1,n)}\left(nD\right)\\
		E_{x}^{(1,n)}\left(nD\right)
	\end{array}\right)=A_{+}\left(\begin{array}{c}
		h_{y}^{(+)}\\
		e_{x}^{(+)}
	\end{array}\right)\exp\left(iq_{+}nD\right),\thinspace n\ge0.
\end{eqnarray*}
Such a representation avoids the appearance of waves exponentially growing
towards $z=\pm\infty$. Substituting these equations into boundary
conditions across the TMD layer,
\begin{eqnarray}
\left(\begin{array}{c}
H_{y}^{(1,0)}\left(0\right)\\
E_{x}^{(1,0)}\left(0\right)
\end{array}\right)=\hat{B}\left(\begin{array}{c}
H_{y}^{(2,-1)}\left(0\right)\\
E_{x}^{(2,-1)}\left(0\right)
\end{array}\right),\label{eq:bc}
\end{eqnarray}
where 
\begin{eqnarray*}
	\hat{B}=\left(\begin{array}{cc}
		1 & -\frac{4\pi}{c}\sigma_{TMD}\left(\omega\right)\\
		0 & 1
	\end{array}\right)
\end{eqnarray*}
is the boundary condition matrix, it is possible to obtain the following (implicit) dispersion relation
\begin{equation}
\frac{h_{y}^{(+)}}{e_{x}^{(+)}}-\frac{h_{y}^{(-)}}{e_{x}^{(-)}}+\frac{4\pi}{c}\sigma_{TMD}\left(\omega\right)=0
\label {disp-rel}
\end{equation}
for the localized eigenstate supported by TMD embedded into the PC (compare to Eq. (\ref{disprel1})). 

The spectrum of the perfectly periodic PC {[}see Fig.\,\ref{fig:infinite_PC_TE}(d){]}
contains a low-frequency stop-band, which vanishes at $k_{y}=0$ (we shall call it \textquotedblleft zeroth-order\textquotedblright), and higher order stop-bands whose width remains finite at $k_{y}=0$.
This fact is crucial for the existence of two different types of localized
eigenstates supported by the inserted TMD layer, whose spectra are shown in Fig.\,\ref{fig:loc-state-TE}(a).
For the particular parameters of Fig.\,\ref{fig:loc-state-TE}(a),
the spectrum contains four distinct modes: (i) two (type A and type B)
in the zeroth stop-band, and (ii) two (type A' and type B') within the
first stop-band. The properties of the A and B modes are similar
to those of the surface EPs supported by TMD cladded by two semi-infinite dielectrics, described by Eq. (\ref{disprel1}) and briefly discussed in the Introduction {[}see Fig.\,\ref{fig:disp-rel-semiinfinite}(b){]}. Yet, there is one distinctive feature: in the PC, the A and B modes bifurcate
from the edge of the first allowed band and can exist on the left of the less steep 
light line, $\omega=ck_{y}/\sqrt{\mathrm{max}\left(\varepsilon_{1},\varepsilon_{2}\right)}$.

The modes inside the first stop-band, named A' and B', have remarkably different properties. They do not approach asymptotically the exciton transition frequencies but rather cross them. 
Moreover, for large $k_{y}$ they occur in the frequency
ranges where the imaginary part of TMD's conductivity is \emph {positive} (shadowed in  Fig.\,\ref{fig:loc-state-TE}(a)). In the frequency range far from $\omega_{A}$ and $\omega_{B}$, ${\rm Im }\sigma_{TMD}$ is small and, as a result, the localized eigenstate appears close to the edge of the allowed band of the PC. At the same time, the spatial profile is rather weakly localized in the vicinity of the TMD layer (actually, the same happens to the "usual" surface EPs; examples are shown in Fig.\,\ref{fig:loc-state-TE}(b) and \ref{fig:loc-state-TE}(d)
	for the type B' and type A modes, respectively). In contrast, for the frequencies near $\omega_{A}$ and $\omega_{B}$, the absolute value of TMD's conductivity is high. As a consequence, the localized eigenstate lies deeply inside the stop-band and its spatial profile is strongly localized {[}see Fig.\,\ref{fig:loc-state-TE}(c){]}. 

	The A' and B' modes, lying within the gap of the PC crystal spectrum, possess another interesting property. Their dispersion curves lie (partially) on the left of the steepest of the two light lines of the dielectrics constituting the PC. Both facts are characteristic of the lossless optical Tamm states (OTS), first described for a heterostructure of two semi-infinite PCs \cite{Kavokin2005}. If the two halves of the structure were identical, there would be no OTS unless the full translation symmetry is broken in some other way like introducing the TMD layer. It also leads to the coupling between the optical mode and the exciton and their anti-crossing as described for a conventional quantum well \cite{Symonds2009} and also for 2D semiconductors in a structure where one of the BRs was replaced by a metallic mirror \cite{Lundt2016}. The "primed" modes of Fig.\,\ref{fig:loc-state-TE} are weakly localized OTS-type modes with \emph {real} Bloch wavevector and the frequency in the vicinity of the PC band edge. They become increasingly excitonic in nature as $k_y$ increases. We shall demonstrate it further in the next section.

\section{Diffraction of light on a TMD-intercalated  PC}
\begin{figure}[t!]
	\centering
	\fbox{
		\includegraphics[width=10cm]{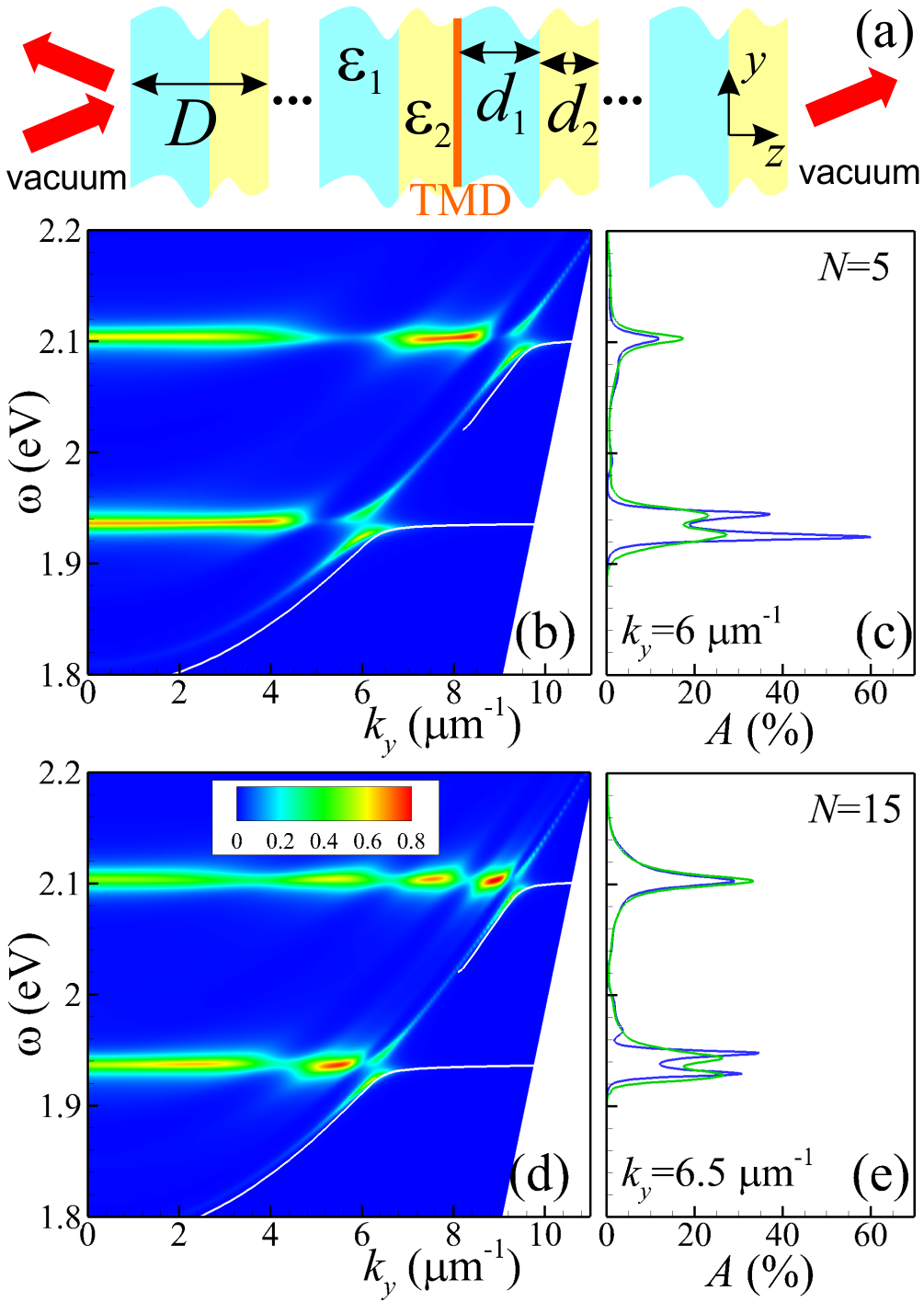}
	}
	\caption{(a) Schematics of a truncated PC, intercalated with a TMD layer.
		The arrows show the incident, reflected and transmitted beams; (b, d) Absorbance (depicted by color map) \emph{versus } frequency and in-plane wavevector $k_{y}=(\omega/c)\sin\theta$ for the structure shown in (a),  containing $N=5$ (b) or $N=15$ (d) unit cells. Other parameters of panel
		(b) are the same as in Figs.\,\ref{fig:infinite_PC_TE} and \ref{fig:loc-state-TE},
		while for panel (d) $\varepsilon_{1}=2.13$, $d_{1}=80\thinspace$nm,
		$\varepsilon_{2}=4$, $d_{2}=130\thinspace$nm. Dispersion curves
		for localized eigenstates in infinite crystal with the same parameters are depicted by white lines. In all cases $\gamma _A=\gamma _B= 4\,$meV. Panels (c) and (e) show the absorbance plotted against $\omega $ for fixed values of $k_y [6\,\mu$m for $N=5$ in (c) and $k_y=6.5\,\mu$m in (e)], for truncated perfect PC (blue lines) and for the structure with layers' thickness disorder (green lines). Relative dispersion of layers' thicknesses for the latter is $\sigma_1/\overline{d_1}=\sigma_2/\overline{d_2}=0.025$.}
	\label{fig:excitation}
\end{figure}

	As discussed above, the localized eigenstates A' and B' lie on the left of the line $\omega=ck_{y}/\sqrt{\mathrm{min}\left(\varepsilon_{1},\varepsilon_{2}\right)}$ and even the vacuum light line in some frequency range. It means that these modes can be coupled to external propagating EM waves, i.e. excited directly by light without using a prism or a grating. 
Let us consider a PC with a finite (and relatively small) number of periods, hosting the TMD layer.
An example of such a structure is shown in Fig.\,\ref{fig:excitation}(a) where
the TMD layer is embedded into a truncated PC containing $2N$ elementary cells intercalated with a TMD layer ($N$ cells before and $N$ cells after the TMD layer). 
Light with frequency $\omega$ falls on the surface of the truncated PC at an angle of incidence $\theta$, which determines the transverse wavevector component $k_{y}$. 

The amplitudes of the incident, $E_{x}^{(i)}$, reflected, $E_{x}^{(r)}$, and transmitted, $E_{x}^{(t)}$, waves in such a truncated PC with the TMD layer inside can be related via the transfer-matrix of the whole structure, $\hat{T}_{tot}$ (see Supplemental Document for details), namely: 
\begin{eqnarray}
&\left(\begin{array}{c}
E_{x}^{(t)}\\
0
\end{array}\right)=\hat{T}_{tot}\left(\begin{array}{c}
E_{x}^{(i)}\\
E_{x}^{(r)}
\end{array}\right)\, .
\label{eq:HE-fields_irt}
\end{eqnarray}
From this equation, the amplitudes of the reflected and transmitted waves can be expressed through the matrix elements of $\hat{T}_{tot}$ as follows:
\begin{eqnarray}
&E_{x}^{(r)}=-\frac{\left(\hat{T}_{tot}\right)_{21}}{\left(\hat{T}_{tot}\right)_{22}}E_{x}^{(i)},\qquad
&E_{x}^{(t)}=\frac{1}{\left(\hat{T}_{tot}\right)_{22}}E_{x}^{(i)}.
\end{eqnarray}

The absorbance ($A$) of the structure can be calculated as the difference between the Poynting vectors'  $z$-components corresponding to the incident, reflected and transmitted waves,
\begin{eqnarray}
A=1-\frac{\left|\left(\hat{T}_{tot}\right)_{21}\right|^2+1}{\left|\left(\hat{T}_{tot}\right)_{22}\right|^2},
\end{eqnarray}
and is depicted in Figs.\,\ref{fig:excitation}(b) and \ref{fig:excitation}(c).
In these plots, the maximal absorption points on the $\left(\omega,\,k_{y}\right)$ plane reveal the characteristic anti-crossings between the horizontal lines representing the A and B excitons and the OTS dispersion curve (accompanying the edge of the PC stop-band). They coincide with the dispersion curves of the A' and B' modes shown by white lines. The small discrepancy between them can be accounted for the photonic crystal truncation (compare Figs.\,\,\ref{fig:excitation}(c) and \ref{fig:excitation}(b), calculated for different $N$). The intensity of the excitonic absorption on the left of the avoided crossing point is modulated also due to the small number of periods in the PCs.

The localized eigenstates are clearly observed in the absorption spectra and can be probed by means of angle-resolved spectroscopy \cite{Lundt2016,Symonds2009} if the frequency and transverse wavevector matching conditions are fulfilled. When the incident light couples to the A' or B' eigenmodes, the incident energy is transferred into the latter and finally dissipated through the exciton in the TMD layer. 

How robust are the reported phenomena against the factors that differ real-world things from theoretical models, such as e.g. fluctuations of the PC-constituting dielectric layers? We addressed this question by calculating the absorbance of TMD-intercalated PCs with Gaussian-distributed random thicknesses of the layers, $d_1$ and $d_2$, with mean values $\overline{d_1}$ and $\overline{d_2}$ as for the perfect structure and standard deviations $\sigma_1$ and $\sigma_2$ (see Sec. 2A of Supplemental Document for details). These results are shown in 
	Figs.\,\ref{fig:excitation}(c) and \ref{fig:excitation}(e), demonstrating that the absorbance of the disordered structure (green lines) is decreased compared to the truncated perfect PC (blue lines), for the A' mode. At the same time, the modes' anti-crossing is still visible for $\sigma_1/\overline{d_1} = \sigma_2/\overline{d_2}$=2.5\%.
	In the Supplemental Document, we present maps similar to panels (b) and (d) of  Fig.\,\ref{fig:excitation}, calculated for stronger disorder ($\sigma /\overline{d}$=10\%), which show that the avoided crossing cannot be resolved anymore. 
	We also notice that it cannot be resolved for the B' mode already with small fluctuations of layers' thicknesses. Somewhat unexpectedly, the intensity of this mode is enhanced by the disorder.   

\section{Conclusions}
We predicted theoretically the existence of two types of localized exciton-polariton eigenstates in 1D photonic crystal with embedded TMD layer. If the excitonic transition frequencies of the TMD layer lie within the first stop-band of the PC for zero transverse wavevector, the spectrum of localized EP states consists of: (i) two (A and B) modes
in the zeroth-order stop-band, and (ii) two (A' and B') in the
first stop-band. The properties of the modes (i) are similar
to those of the surface EPs supported by a single TMD cladded by two semi-infinite dielectrics, while modes (ii) are related to the optical Tamm states \cite{Kavokin2005,TammPlasmonTP,Symonds2009}, although the analogy is only partial in both cases. 

	The A' and B' modes result from the coupling of the TMD exciton to the Tamm-type state of light in the photonic crystal with the translational symmetry broken by the inserted TMD layer. Coupling of such a photonic state localized by two reflectors to elementary excitations in a 2D material has recently been described for the case of graphene plasmons \cite{Silva2019}. 
The Tamm-type modes predicted here can be effectively coupled to propagating light waves falling directly on the surface of the PC with a relatively small number of unit cells. In the presented simulated example {[}Fig.\,\ref{fig:excitation}(c){]} we demonstrated the feasibility of excitation of such localized eigenstate in an experimentally attainable structure with the dielectric constants $\varepsilon_{1}=2.13$ and $\varepsilon_{2}=4.04$, corresponding to a SiO$_{2}$/Si$_{3}$N$_{4}$ BR used in Ref. \cite{Sanchez2016} (these and other inexpensive dielectrics are commonly used for making reasonable Bragg reflectors \cite{Yepuri2020-BRs}).
Methods similar to the suggested experiment were already used \cite{sasin_2010,Symonds2009,Sanchez2016} for demonstration of optical Tamm states, even though one of the BRs was replaced by another type of mirror, such as a metal or an organic dye film. Our simulations for the structure with natural disorder (fluctuations of layers' thicknesses) demonstrate that the characteristic avoided crossing can be observed for the A' mode if the relative dispersion of the thicknesses is below $\approx$\,5\%. Therefore, we hope that this article may stimulate experiments aimed at observing this new type of localized exciton-polaritons. 

\section*{Acknowledgements}
Y.V.B., N.M.R.P., and M.I.V. acknowledge support from the European
Commission through the project \textquotedblleft Graphene-Driven Revolutions
in ICT and Beyond\textquotedblright - Core 3 (Ref. No. 881603) and the Portuguese
Foundation for Science and Technology (FCT) in the framework of the Strategic Funding UIDB/04650/2020. Authors also acknowledge FEDER
and FCT for support  through
projects POCI-01-0145-FEDER-028114 and PTDC/FIS-MAC/28887/2017. The authors declare no conflicts of interest.

\bibliographystyle{unsrtnat}
\bibliography{tmd_phot_crys_bib}  

\section{Localized polariton states in a photonic crystal intercalated by a transition metal dichalcogenide monolayer: supplemental document}
	\subsection{Transfer matrix formalism for a perfect PC}
	
	The solutions of the Maxwell's equations in each slab can be expressed via transfer-matrices, 
	\begin{eqnarray}
	\hat{T}_{m}\left(z\right)=\left(\begin{array}{cc}
	\cos\left(k_{z}^{(m)}z\right) & i\frac{k_{z}^{(m)}}{\kappa}\sin\left(k_{z}^{(m)}z\right)\\
	i\frac{\kappa}{k_{z}^{(m)}}\sin\left(k_{z}^{(m)}z\right) & \cos\left(k_{z}^{(m)}z\right)\;,
	\end{array}\right)\label{eq:transfer-matrix}
	\end{eqnarray}
	as
	\begin{eqnarray*}
		&\left(\begin{array}{c}
			H_{y}^{(1,n)}\left(z\right)\\
			E_{x}^{(1,n)}\left(z\right)
		\end{array}\right)=\hat{T}_{1}\left(z-nD\right)\left(\begin{array}{c}
			H_{y}^{(1,n)}\left(nD\right)\\
			E_{x}^{(1,n)}\left(nD\right)
		\end{array}\right),\\
		&\left(\begin{array}{c}
			H_{y}^{(2,n)}\left(z\right)\\
			E_{x}^{(2,n)}\left(z\right)
		\end{array}\right)=\hat{T}_{2}\left(z-nD-d_{1}\right)\left(\begin{array}{c}
			H_{y}^{(2,n)}\left(nD+d_{1}\right)\\
			E_{x}^{(2,n)}\left(nD+d_{1}\right)
		\end{array}\right),
	\end{eqnarray*}
	for spatial domains $nD<z<nD+d_{1}$ and $nD+d_{1}<z<(n+1)D$, respectively.
	The tangential components of the field across the PC's period
	can be related via the product of the transfer-matrices, $\hat{T}_{12}=\hat{T}_{2}\left(d_{2}\right)\hat{T}_{1}\left(d_{1}\right)$, Eq. (5) of the main text, which expresses the continuity of the tangential components of the field across the dielectrics' boundaries. 
	
	Applying the Bloch theorem, 
	$$\left(H_{y}^{(1,n)}\left(nD\right)\quad E_{x}^{(1,n)}\left(nD\right)\right)^{T}=\left(H_{y}^{(1,0)}\left(0\right)\quad E_{x}^{(1,0)}\left(0\right)\right)^{T}\exp\left(iqnD\right)\, ,$$
	we obtain: 
	\begin{eqnarray}
	\left[\hat{T}_{12}-\hat{I}\exp\left(iqD\right)\right]\left(\begin{array}{c}
	H_{y}^{(1,0)}\left(0\right)\\
	E_{x}^{(1,0)}\left(0\right)
	\end{array}\right)=0\, ,
	\label{eq:HE-Bloch}
	\end{eqnarray}
	where $\hat{I}$ is the 2$\times$2 unity matrix and $q$ stands for the 
	Bloch wavevector. 
	
	The dispersion relation of electromagnetic waves inside the PC, as well as the 
	general solution of Eq.\,(\ref{eq:HE-Bloch}), can be represented
	in terms of eigenvalues, $\lambda_{\pm}$, and eigenvectors,
	$\left(h_{y}^{(\pm)}\quad e_{x}^{(\pm)}\right)^{T}$, of the matrix $\hat{T}_{12}$.
	This relation, coming from the compatibility condition of Eq.\,(\ref{eq:HE-Bloch}), can be expressed as Eq.(6) of the main text. { Notice that the matrix $\hat{T}_{12}$ is unimodular, since so are the matrices $\hat{T}_{m}\left(z\right)$.}
	
	Let us notice that it is possible to obtain a general
	expression for the EM field components at a distance from the $z=0$ plane equal to an integer number of periods as a superposition of forward- and
	backward-propagating waves, using the Bloch theorem and the eigenvectors,
	\begin{eqnarray}
	\left(\begin{array}{c}
	H_{y}^{(1,n)}\left(nD\right)\\
	E_{x}^{(1,n)}\left(nD\right)
	\end{array}\right)=\sum_{j=\pm}A_{j}\left(\begin{array}{c}
	h_{y}^{(j)}\\
	e_{x}^{(j)}
	\end{array}\right)\exp\left(iq_{j}nD\right),
	\end{eqnarray}
	where $A_{\pm}$ are the amplitudes of the respective waves. This relation is used in the main text. 
	
	\subsection{Transmission of an EM wave through a truncated photonic crystal}
	\subsubsection{Truncated perfect PC}
	To calculate the absorbance of a finite PC (consisting of $2N$ unit cells) with an embedded TMD layer in the middle, shown in Fig.\,4(a) of the main text, we apply sequentially Eqs.\,(5) and (7) of the main text and obtain the following relation between the fields at PC's boundaries: 
	\begin{eqnarray}
	&\left(\begin{array}{c}
	H_{y}^{(2,N)}\left(ND\right)\\
	E_{x}^{(2,N)}\left(ND\right)
	\end{array}\right)=\left(\hat{T}_{12}\right)^N\hat{B}\left(\hat{T}_{12}\right)^N\left(\begin{array}{c}
	H_{y}^{(1,-N)}\left(-ND\right)\\
	E_{x}^{(1,-N)}\left(-ND\right)
	\end{array}\right).\label{eq:HE-boundaries}
	\end{eqnarray}
	Outside of the truncated PC (in the regions with $\varepsilon_m=1$) the solutions of Maxwell's equations [(3) and (4) of the main text] can be represented in the matrix form as
	\begin{eqnarray}
	&\left(\begin{array}{c}
	H_{y}^{(-N-1)}\left(z\right)\\
	E_{x}^{(-N-1)}\left(z\right)
	\end{array}\right)=\hat{F}\left(\begin{array}{c}
	E_{x}^{(i)}\exp\left[ik_z\left(z+ND\right)\right]\\
	E_{x}^{(r)}\exp\left[-ik_z\left(z+ND\right)\right]
	\end{array}\right),\label{eq:fields-ie-vacuum}\\
	&\left(\begin{array}{c}
	H_{y}^{(N+1)}\left(z\right)\\
	E_{x}^{(N+1)}\left(z\right)
	\end{array}\right)=\hat{F}\left(\begin{array}{c}
	E_{x}^{(t)}\exp\left[ik_z\left(z-ND\right)\right]\\
	0
	\end{array}\right).\label{eq:fields-t-vacuum}
	\end{eqnarray}
	Here the superscripts $(-N-1)$ and ($N+1$) correspond to the electromagnetic fields in the spatial domain $z<-ND$ and $z>ND$, respectively; $E_{x}^{(i)}$, $E_{x}^{(r)}$, and $E_{x}^{(t)}$ stand for the amplitudes of the electric field's $x$-component of the incident, reflected, and  transmitted waves, respectively, $k_z=\sqrt{\kappa^2-k_y^2}$ is $z$-component of the wavevector in vacuum, and
	\begin{eqnarray}
	&\hat{F}=\left(\begin{array}{cc}
	k_z\kappa^{-1} & -k_z\kappa^{-1}\\
	1 & 1
	\end{array}\right)
	\end{eqnarray}
	is the field matrix. In Eqs.\,(\ref{eq:fields-ie-vacuum})--(\ref{eq:fields-t-vacuum}), the incident and transmitted waves are assumed to be forward-propagating, while the reflected wave is backward-propagating. Using the condition of continuity of the tangential components at boundaries $z=\pm ND$, it is possible to obtain a relation between the amplitudes of the incident, transmitted, and reflected waves from Eqs.\,(\ref{eq:HE-boundaries}), (\ref{eq:fields-ie-vacuum}), and (\ref{eq:fields-t-vacuum}), which yields the absorbance, Eq. (11) of the main text, {where \begin{eqnarray}\hat{T}_{tot}=\left(\hat{F}\right)^{-1}\left(\hat{T}_{12}\right)^N\hat{B}\left(\hat{T}_{12}\right)^N\hat{F}
		\label{eq:Ttotal}
		\end{eqnarray} 
		is the transfer-matrix of the whole structure.
	}
	
	\subsubsection{Impact of disorder}
	Let us now consider the wave transmission through a truncated PC where the thicknesses of the layers are random values. We shall assume that the probability density of the thickness $d_m$ ($m=1,2$) follows a Gaussian distribution of the form
	\begin{eqnarray}
	p\left(d_m\right)=\frac{1}{\sigma_m\sqrt{2\pi}}\exp\left[-\frac{1}{2}\left(\frac{d_m-\overline{d_m}}{\sigma_m}\right)^2\right]\,,
	\label{eq:Gaussian}
	\end{eqnarray} 
	where $\overline{d_m}$ and $\sigma_m$ stand for the mean value and the standard deviation, respectively. 
	In order to calculate the $l$-th realization of the absorbance, $A^{(l)}$, $2\times 2N$  random numbers, $d_m^{(l,n)}$, obeying the distribution (\ref{eq:Gaussian}) were generated ($n=1,\dots,2N$), which were used to obtain the $l$-th realization of the transfer-matrix of the whole structure:
	\begin{eqnarray}
	\hat{T}_{tot}^{(l)}=\left(\hat{F}\right)^{-1}\prod_{n=0}^{N-1}\left[\hat{T}_{2}\left(d_{2}^{(l,n)}\right)\hat{T}_{1}\left(d_{1}^{(l,n)}\right)\right]\hat{B}\nonumber\\
	\times\prod_{n=N}^{2N-1}\left[\hat{T}_{2}\left(d_{2}^{(l,n)}\right)\hat{T}_{1}\left(d_{1}^{(l,n)}\right)\right]\hat{F}\, .
	\label{eq:Ttotal-disordered}
	\end{eqnarray} 
	This matrix (\ref {eq:Ttotal-disordered}) was inserted into Eq.\,(11) of the main text,
	\begin{eqnarray}
	A^{(l)}=1-\frac{\left|\left(\hat{T^{(l)}}_{tot}\right)_{21}\right|^2+1}{\left|\left(\hat{T^{(l)}}_{tot}\right)_{22}\right|^2}\, .
	\end{eqnarray} 
	The average absorbance, $\overline{A}$, of the disordered structure was calculated as
	\begin{eqnarray}
	\overline{A}=M^{-1}\sum_{l=0}^{M-1}A^{(l)},
	\end{eqnarray}
	where $M$ is the number of absorbance realizations. 
	\begin{figure}[t!]
		\centering
		\fbox{
			\includegraphics[width=10cm]{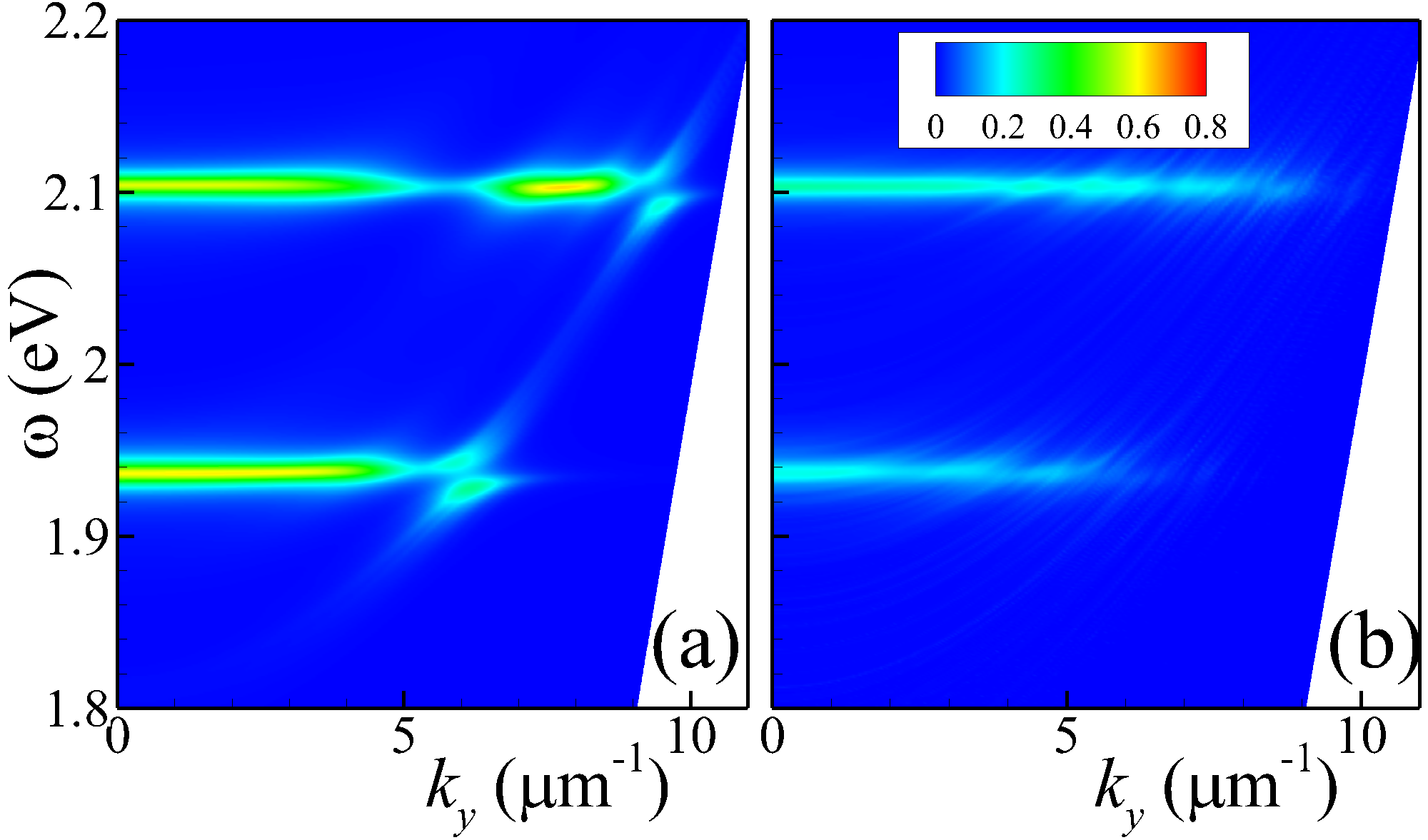}
		}
		\caption{Average absorbance $\overline{A}$ (depicted by color map) \emph{versus } frequency and in-plane wavevector $k_{y}=(\omega/c)\sin\theta$ for the disordered structure with the following parameters: $N=5$, $\varepsilon_{1}=2$,
			$\overline{d_1}=140\thinspace$nm, $\varepsilon_{2}=12$, $\overline{d_2}=70\thinspace$nm, $\sigma_1/\overline{d_1}=\sigma_2/\overline{d_2}=0.025$ (a), or $\sigma_1/\overline{d_1}=\sigma_2/\overline{d_2}=0.1$ (b).  }
		\label{fig:excitation_sup}
	\end{figure}
	The results are shown in Fig.\,\ref{fig:excitation_sup} for two magnitudes of disorder. As seen from the figure, the mode anti-crossing is still visible for $\sigma /\overline d$=2.5\%, although the PC band edge is less sharp than in the case of truncated perfect PC [see Fig.\,4(b) in the main text] owing the presence of disorder. At the same time, the anti-crossing disappears completely in the case of strong disorder, $\sigma /\overline d$=10\% [Fig.\,\ref{fig:excitation_sup}(b)].






\end{document}